\documentclass[prb,aps,twocolumn,superscriptaddress,nofootinbib,longbibliography,notitlepage]{revtex4-1} %---APS---SI---arxiv with ref. titles %---use pra for SI
\usepackage{amsmath} \usepackage{amssymb} \usepackage{amsfonts}
\usepackage{mathrsfs} \usepackage{mathtools} \usepackage{bm} \usepackage{bbm} \usepackage{braket} \usepackage{color} \usepackage{comment} \usepackage{dcolumn} \usepackage{enumerate} \usepackage{epsfig} \usepackage{gensymb} \usepackage{graphicx} \usepackage{indentfirst} \usepackage{lmodern} \usepackage{psfrag} \usepackage{pst-all} \usepackage{soul} \usepackage{xcolor} %\usepackage{units}
\usepackage{float} %---APS---SI---arxiv
\usepackage{cancel}
\usepackage{verbatim}

\usepackage[colorlinks,linkcolor=blue,citecolor=blue,urlcolor=blue,hyperindex,driverfallback=dvipdfm]{hyperref}

 \usepackage[T1]{fontenc} %---APS---ACS---SI---arxiv
\def\ii{{\rm i}} \def\ee{{\rm e}}
\def\me{m_{\rm e}} 
\def\Ree{{\rm Re}} \def\Imm{{\rm Im}}
                          \def\vb{{\bf v}} %--- bold vectors
 \def\yy{\hat{\bf y}} \def\zz{\hat{\bf z}}     
\def\kpar{k_\parallel}  %--- unit vectors 
\def\rp{r_{\rm p}} \def\rs{r_{\rm s}} %--- Fresnel's coefficients
\def\e{{\rm e}}

\begin{document} %---APS---SI---arxiv
% =========================================================
% --- title, affiliations, abstract -----------------------
% =========================================================
\title{Unity-order coupling between free electrons and multiphoton waveguided Fock states
%Tunable free-electron coupling to guided modes via electrostatic trajectory control
}

% --- OSA affiliations ------------------------------------
%\author[1,2,*]{F.~Javier~Garc\'{\i}a~de~Abajo}
%\affil[1]{ICFO-Institut de Ciencies Fotoniques, The Barcelona Institute of Science and Technology, 08860 Castelldefels (Barcelona), Spain}
%\affil[2]{ICREA-Instituci\'o Catalana de Recerca i Estudis Avan\c{c}ats, Passeig Llu\'{\i}s Companys 23, 08010 Barcelona, Spain}
%\affil[*]{E-mail: javier.garciadeabajo@nanophotonics.es}

% --- APS,SI,arxiv affiliations ---------------------------
 
\author{Leila~Prelat}
\author{Saad~Abdullah}
\author{Cruz~I.~Velasco}
\affiliation{ICFO-Institut de Ciencies Fotoniques, The Barcelona Institute of Science and Technology, 08860 Castelldefels (Barcelona), Spain}
\author{F.~Javier~Garc\'{\i}a~de~Abajo}
\email{javier.garciadeabajo@nanophotonics.es}
\affiliation{ICFO-Institut de Ciencies Fotoniques, The Barcelona Institute of Science and Technology, 08860 Castelldefels (Barcelona), Spain}
\affiliation{ICREA-Instituci\'o Catalana de Recerca i Estudis Avan\c{c}ats, Passeig Llu\'{\i}s Companys 23, 08010 Barcelona, Spain}
%\email{javier.garciadeabajo@nanophotonics.es}
%\affiliation{ICFO-Institut de Ciencies Fotoniques, The Barcelona Institute of Science and Technology, 08860 Castelldefels (Barcelona), Spain}
%\affiliation{ICREA-Instituci\'o Catalana de Recerca i Estudis Avan\c{c}ats, Passeig Llu\'{\i}s Companys 23, 08010 Barcelona, Spain}

% --- document format -------------------------------------
%\begin{document} %---ACS
%\ociscodes{XXX} %(300.6530) Ultrafast spectroscopy; (270.1670) Coherent optical effects.} %---OSA

% --- abstract --------------------------------------------
\begin{abstract}
Electron beams enable highly localized near-field excitation of waveguided optical modes, yet their coupling is typically limited by short interaction times along straight-line trajectories with fixed impact parameters. Here, we theoretically demonstrate that electrostatic steering overcomes this limitation by introducing a tunable turning point in grazing electron trajectories, thus controlling the minimum electron--waveguide separation and producing strong coupling to waveguided modes. Specifically, we consider a biased rectangular silicon waveguide, where a repulsive static field deflects a grazing electron. In this configuration, the electron turning point governs both the coupling strength and the modal selectivity, which can be dynamically tuned through the electron incidence angle and the applied bias. In addition, the aloof electron--waveguide interaction suppresses lossy high-energy channels (e.g., above the silicon band gap) while preserving substantial excitation of the targeted waveguided modes. Using a practical biasing configuration and 100~keV electrons, we predict an average yield exceeding ten photons per electron, with voltage-tunable control of the interaction. Our results establish electrostatic steering as a practical route for engineering and enhancing free-electron coupling to waveguided photonic modes.
\end{abstract}

% --- document format -------------------------------------
%\setboolean{displaycopyright}{true} %---OSA
%\begin{document} %---OSA
\maketitle %---APS---OSA---SI---arxiv
\date{\today} %---APS---arxiv
%\tableofcontents %---APS---SI---arxiv optional
%\setcounter{equation}{0} %---OSA
%\setkeys{acs}{maxauthors=0} %---ACS to avoid "et al." in references
%\noindent \textbf{Keywords:} nanophotonics, free electrons, guided modes, tunable electron trajectory %---ACS

%\vfill
% =========================================================
\section{Introduction}

Optical dielectric waveguides constitute a fundamental architecture for integrated photonics,\cite{SBC24,BC18,PGS24} enabling low-loss routing,\cite{JOGM23,CWB22} spatial confinement, and coherent on-chip manipulation of electromagnetic fields.\cite{MCL21} In particular, Si waveguides provide strong electromagnetic confinement and low-loss propagation in the near-infrared,\cite{PTK24} as evidenced by the dispersion relations and evanescent near-field profiles that characterize their guided modes.\cite{SL12,M13_1} In many applications, it is desirable to excite selected modes while controlling the spectral distribution of the coupled field. 

As an alternative to conventional optical coupling schemes based on gratings or edge-coupling geometries, electron beams (e-beams) have emerged as a powerful platform for the localized excitation and probing of guided modes.\cite{KAD21,RCKY23,BBV16,CVG13,paper180,paper430,paper123} Free electrons were postulated as an efficient tool to generate heralded single photons.\cite{paper180} Recent work has materialized this proposal with the experimental generation of electron-heralded nonclassical light in an integrated platform,\cite{AHF25} including multiphoton number Fock states. The evanescent field of a swift electron carries large in-plane momenta,\cite{paper149,CKY23} providing access to near-field mode profiles that are difficult to reach with far-field illumination.\cite{DMP22,AFD21,LTA13_2} Furthermore, e-beams constitute a minimally invasive probe with high spatial resolution in electron microscopes, as they can be laterally focused down to sub-\AA{}ngstr\"om dimensions.\cite{MKM08,NCD04,proc058} Their interaction with the electromagnetic environment of a specimen can be quantitatively resolved through electron energy-loss spectroscopy (EELS). By measuring the inelastic scattering probability as a function of energy and momentum transfer,\cite{PSV1975,E96} EELS facilitates the mapping of optical and polaritonic excitations created in the specimen with meV-scale energy resolution in state-of-the-art instruments.\cite{KLD14,HRK20}
 
The coupling strength to waveguide modes depends sensitively on the electron–waveguide separation. Tuning the minimum approach distance, therefore, provides a practical means of favoring low-loss guided channels while suppressing the relative contribution of high-energy absorptive processes. However, achieving modal and spectral selectivity remains challenging in standard electron-driven excitation schemes. In conventional configurations, the coupling between an electron and a photonic structure is largely determined by the electron kinetic energy and a fixed impact parameter.\cite{paper433,paper432} As a result, multiple degenerate or nearly degenerate modes can be excited simultaneously, together with undesired lossy, high-energy channels, such as those associated with valence-electron promotion above the band gap of dielectric materials.\cite{PGS24,AS1983} Experiments clearly show that the electron-incidence geometry can influence the optical response of Si.\cite{CYM10} Nevertheless, although trajectory engineering has been proposed as a theoretical control parameter,\cite{paper455} its practical implementation for mode-selective excitation in integrated platforms remains largely unexplored.

In this work, we demonstrate that shaping the electron trajectory through external DC fields provides a robust mechanism for achieving mode selectivity and enhancing the photon yield per electron. For illustration, we investigate the interaction between an e-beam and a rectangular Si waveguide. We first analyze the electromagnetic field produced by a passing electron and identify the poles of the waveguide optical response to determine its supported modes. We then calculate the EELS probability for an electron traveling parallel to an extended planar waveguide (Sec.~\ref{sec_parallel_trajectory}), and an electron following a grazing-incidence trajectory along a one-dimensional (1D) waveguide and being deflected by an applied electrostatic potential (Sec.~\ref{sec_parabolic_trajectory}). The curved electron path exhibits a tunable turning point corresponding to the minimum electron--waveguide separation, the position of which can be controlled through the incidence angle and the applied electrostatic bias, and it plays a central role in determining both the coupling strength and modal selectivity.

For the deflected trajectory, we first consider a tutorial example with a uniform external DC field that produces a parabolic path, generated by a differential potential applied between the waveguide and an electrode placed above it (vertical gating). We then discuss a more practical scenario involving a spatially varying external DC field for a configuration involving a doped Si waveguide on a sapphire substrate flanked by two waveguide-like Si linear gates (lateral gating). In the latter configuration, we obtain a substantial photon yield ($>10$ photons per electron) with experimentally attainable geometrical, electrical, and e-beam parameters. Our analysis accounts for the dynamics of the electron near the dielectric interface, including image-potential effects.

% --- Figure 1 --------------------------------------------
\begin{figure}[htbp]
\centering
\includegraphics[width=\linewidth]{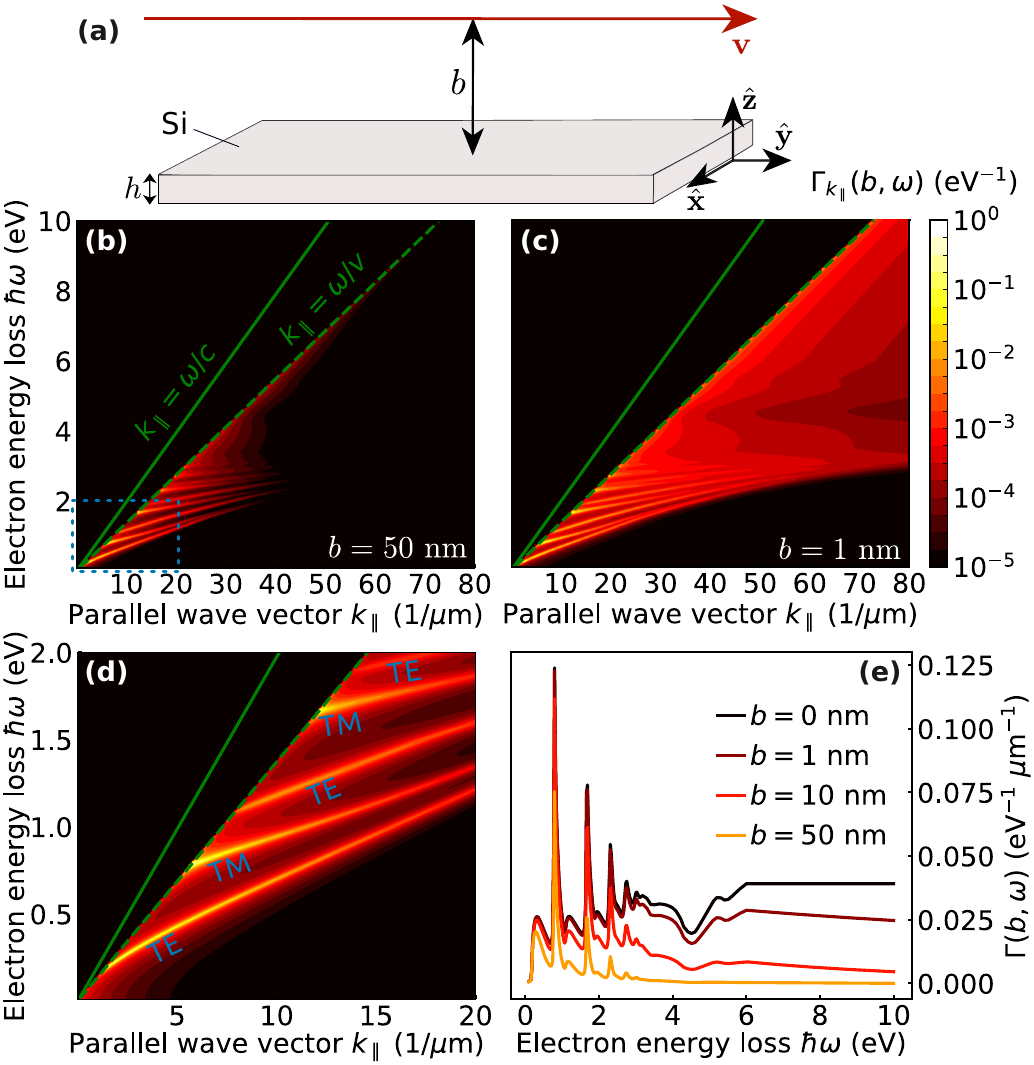}
\caption{\textbf{Free-electron coupling to a planar waveguide in a parallel-trajectory configuration.}
\textbf{(a)}~Schematic representation of an e-beam moving with velocity $\vb=v\yy$ at a distance $b$ from a self-standing laterally extended Si waveguide of thickness $h$.
\textbf{(b,c)}~Wave-vector- and energy-resolved EELS probability per unit of path length [Eq.~(\ref{eels_b_plane})] for (b) $b=50$~nm and (c) $b=1$~nm.
\textbf{(d)}~Close-up of the low-energy region in (b). Labels indicate the TM and TE nature of the waveguided modes.
\textbf{(e)}~EELS probability per unit of path length [Eq.~(\ref{eels_a_plane})] for different values of $b$.
In all panels, the waveguide thickness is $h=200~$nm, and the electron kinetic energy is $200~$keV. Solid and dashed green lines in (b-d) represent the light and electron lines, respectively.}
\label{Fig1}
\end{figure}

% =========================================================
\section{Electron coupling to a planar silicon waveguide}
\label{sec_parallel_trajectory}

As a reference configuration, we first study the system sketched in Fig.~\ref{Fig1}(a), consisting of a self-standing infinitely extended planar Si waveguide of thickness $h$, and an e-beam moving parallel to the surface with velocity $\vb=v\yy$ at a distance $b$ from the waveguide. Within the local-response approximation adopted in this work, the momentum-resolved EELS probability per unit of path length is given by\cite{paper149}
\begin{subequations} \label{eels_plane} %--
\begin{align} \label{eels_a_plane} %--
&\Gamma(b,\omega) = \int_{\omega/v}^\infty d\kpar \;\Gamma_{\kpar}(b,\omega), \\
&\Gamma_{\kpar}(b,\omega) = \dfrac{2e^2\kappa}{\pi\hbar v^2\kpar k_x}\ee^{-2\kappa b} \label{eels_b_plane} \\
&\quad\quad\quad\times\Imm \Big\{\rp(\kpar,\omega)+\Big(\dfrac{k_x v}{\kappa c}\Big)^2 \rs(\kpar,\omega)\Big\}, \nonumber
\end{align}
\end{subequations}
where we explicitly indicate the dependence on electron--waveguide distance $b$. Here, $e>0$ is the elementary charge, $\kpar$ and $\omega$ are the total in-plane wave vector and the angular frequency of the electromagnetic interaction, $k_x=\sqrt{\kpar^2-\omega^2/v^2}$ is the component of the wave vector along the in-plane $x$ direction (perpendicular to the trajectory), $\kappa=\sqrt{\kpar^2-k^2}$ with $k=\omega/c$ describes the exponential attenuation of the EELS probability with $b$, and $r_\nu$ denotes the Fresnel reflection coefficients of the Si waveguide for $\nu=\,$s and p polarizations. The latter are calculated as $r_\nu= r_\nu^{12}(1- \ee^{-2\kappa'h})[1 - (r_\nu^{12})^2 \ee^{-2\kappa'h}]$, where $\kappa'=\sqrt{\kpar^2-\epsilon k^2-\ii 0^+}$ (with ${\rm Re}\{\kappa'\}>0$) is the out-of-plane component of the optical wave vector in Si (permittivity $\epsilon$ taken from tabulated measured data\cite{AS1983}), and $\rs^{12} = (\kappa-\kappa')/(\kappa+\kappa')$ and $\rp^{12}=(\epsilon\kappa-\kappa')/(\epsilon\kappa+\kappa')$ are the reflection coefficients of a semi-infinite Si surface for incidence from vacuum.

Figures~\ref{Fig1}(b) and (c) show the wave-vector- and energy-resolved electron energy-loss probability per unit of path length calculated from Eq.~(\ref{eels_b_plane}) for two impact parameters, $b = 50$~nm and $b = 1$~nm, respectively. The solid and dashed green curves indicate the light line in vacuum ($\kpar=\omega/c$) and the electron line ($\kpar=\omega/v$), respectively. Modes are excited only below the electron line. Comparing the two impact parameters in Figs.~\ref{Fig1}(b) and (c), we observe that the electron creates a larger number of excitations when it moves closer to the waveguide, and in particular, higher-$\kpar$ modes possess a shorter vacuum penetration length that prevents their excitation for large $b$. Consequently, waveguided modes are the main excitation channels for large $b$, while a continuum of excitations corresponding to the creation of electron-hole pairs in the material dominates the dispersion diagram at small $b$. We are interested in a clean excitation of waveguide modes, which are shown in Fig.~\ref{Fig1}(d) [a close-up of \ref{Fig1}(b)] at low excitation energies, including labels that identify their transverse-magnetic (TM) and transverse-electric (TE) character. These types of modes are signaled by the conditions $\rp=0$ and $\rs=0$, respectively [see Eq.~(\ref{eels_b_plane})].

For a better quantification of the dependence of undesired electron-hole-pair excitations (across the Si band gap) on the electron--surface distance $b$, we show in Fig.~\ref{Fig1}(e) the EELS probability per unit of path length as a function of energy loss for different values of $b$, as obtained by numerically solving Eq.~(\ref{eels_a_plane}). A main conclusion from these results is that waveguided modes [i.e., those identified in Fig.~\ref{Fig1}(d)] can still be efficiently excited for relatively large impact parameters $b$, while losses associated with interband transitions are strongly suppressed. In what follows, we shall be interested in relatively large values of $b$ (tens of nanometers), so that only long-lived waveguide modes are excited, while inelastic interband transitions can be neglected. This regime plays an important role when leveraging entanglement between electron energy losses and photons created in the waveguide to implement enhanced sensing protocols in which electron coherence is critical.\cite{paper455}

% =========================================================
\section{Electron coupling to electrically gated optical waveguides}
\label{sec_parabolic_trajectory}

\begin{figure}[htbp]
\centering
\includegraphics[width=\linewidth]{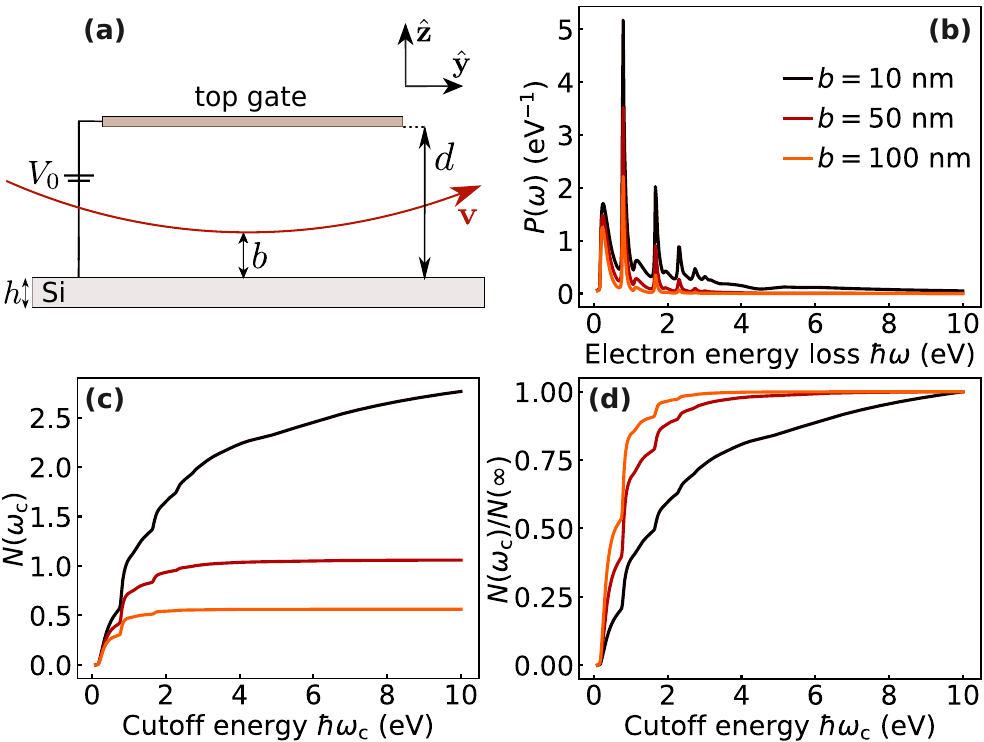}
\caption{\textbf{Electron coupling to a planar waveguide for a parabolic trajectory}.
\textbf{(a)}~Schematic representation of an e-beam under grazing incidence on an extended planar Si waveguide of thickness $h$, separated by a distance $d$ from a top gate biased at a voltage $V_0$. We consider $d$ to be large enough to neglect the electron--waveguide interaction at the point of electron entrance in the capacitor region. The parabolic electron trajectory is determined by the vertical uniform repulsive electric field $E^{\rm DC}=V_0/d>0$ and the minimum distance $b$ from the waveguide (turning point).
\textbf{(b)}~Energy-resolved EELS probability $P(\omega)$ [Eq.~(\ref{eels_oblique})] for different values of $b$.
\textbf{(c)}~Energy-integrated EELS probability $N(\omega_{\rm c})$ as a function of cutoff energy $\hbar\omega_{\rm c}$ [Eq.~(\ref{eq_Gamma_f})] for the same values of $b$ as in (b).
\textbf{(d)}~Same as (c), normalized to the large-cutoff limit $N(\infty)$.
In all panels, the waveguide thickness is $h=200~$nm, the electron energy is $200~$keV, and the DC field is $E^{\rm DC}=10^8\,$V/m.}
\label{Fig2}
\end{figure}

We intend to shape the electron trajectory in such a way that it interacts over prolonged interaction lengths with a waveguide, thus boosting the total photon emission probability per incident electron. To do so, we introduce a DC field that reflects the e-beam at a certain distance of maximum approach to the waveguide [see Figs.~\ref{Fig2}(a) and \ref{Fig3}(a)]. For tutorial purposes, we consider first an extended planar waveguide and a uniform electric field such as that created in a capacitor under a vertical gating configuration (Fig.~\ref{Fig2}). We then move to a more practical design consisting of a gated 1D optical waveguide alongside two adjacent gates running parallel to it, such that a repulsive field is also produced that bounces the electron at a distance $b$ from the waveguide (Fig.~\ref{Fig3}).

We quantify the electron-waveguide coupling by integrating the EELS probability along the electron trajectory, assuming that the velocity component parallel to the waveguide remains approximately constant, while the transverse velocity takes much smaller values. By playing with the gating conditions, we modify the electron trajectory to increase the electron--waveguide interaction length while avoiding electron--surface collisions. 

% ---------------------------------------------------------
\subsection{Electron trajectory in the presence of electrostatic and image forces}
\label{electron_trajectory}

We consider the electron to interact with the gated waveguide under glancing conditions relative to the waveguide surface plane $z=0$. The electron reaches a minimum distance $z=b$ as a result of the repulsive force $F^{\rm DC}_z(z)=-dU^{\rm DC}(z)/dz>0$ produced by an applied DC potential energy $U^{\rm DC}(z)$, which deflects the electron. We take the waveguide to run along the $y$ direction and limit our analysis to electron trajectories contained in the $y-z$ plane, with the $y$ velocity component taken to remain approximately constant in the calculation of the loss probability and the electron trajectory.

The image-charge interaction between the electron and the waveguide surface produces an attractive image force $F^{\mathrm{im}}_z(z)=-dU^{\rm im}(z)/dz<0$, which we incorporate in our calculations through a velocity-dependent image potential approximated as
\begin{align} \label{eq_potential_image_approx} %--
U^{\rm im}(z)=-\frac{e^2}{4\gamma z},
\end{align}
which is obtained by assimilating the waveguide surface to an extended perfect-electric-conductor (PEC) plane. This expression is similar to the classical static image potential, but corrected by a Lorentz factor $\gamma=1/\sqrt{1-v^2/c^2}$ that accounts for the effect of the relativistic electron velocity (see Appendix~\ref{image_force}).

The electron motion along the out-of-plane direction $\zz$ is described by
\begin{align} \label{eq_motion} %--
m_{\rm e}\gamma\,\dfrac{d^2 z }{dt^2}=-\frac{d}{dz}\big[U^{\rm DC}(z)+U^{\rm im}(z)\big], 
\end{align}
where $m_{\rm e}$ is the electron mass. The transverse velocity is assumed to satisfy $v_\perp\ll v$ at all times, while the total energy loss is small compared with the incident electron kinetic energy, and therefore, we can assume that both $v$ and $\gamma$ remain nearly constant during the interaction. Under this approximation, the in-plane electron coordinate evolves as $[x=0,y(t)=vt]$, while the out-of-plane coordinate $z(t)$ follows Eq.~(\ref{eq_motion}). Multiplying by $dz/dt$, the latter can readily be reduced to the first-order ordinary differential equation
$(dz/dt)^2 =(2/\me\gamma)\big[C-U^{\rm DC}(z)-U^{\mathrm{im}}(z)\big]$, where $C$ is an integration constant that we determine by imposing an electron turning point ($dz/dt=0$) at  $z=b$. Using $dy \approx v\, dt$, this equation can be transformed into
\begin{subequations}
\begin{align} \label{eq_dy_electron_aux} %--
&\frac{dy}{dz}=\sqrt{\frac{\me v^2\gamma/2}{U(b)-U(z)}}, \\
&U(z)=U^{\rm DC}(z)+U^{\rm im}(z) \label{gofz} %--
\end{align}
\end{subequations}
for the outgoing part of the trajectory ($y>0$, past the turning point at $y=0$). From this expression, upon integration over $z$, we obtain the electron position $z(y)$ as a function of distance $y$ traveled along the waveguide. Also, the incoming part of the trajectory is determined by the symmetry $z(y)=z(-y)$.

It should be noted that, if the magnitude of the image force exceeds that of the external force at $z=b$, no turning point can be found, and the electron collides with the waveguide. Consequently, a threshold  value $b_{\rm th}$ can be defined by the condition
\begin{align} \label{bthreshold} %--
F_z^{\rm DC}(b_{\rm th})=-F_z^{\rm im}(b_{\rm th}),
\end{align}
such that we must have $b>b_{\rm th}$ to avoid electron--surface impact. Importantly, as we show below (see Appendix~\ref{image_force}), the values of $b_{\rm th}$ derived from Eq.~(\ref{bthreshold}) using the PEC expression for the image potential [Eq.~(\ref{eq_potential_image_approx})] agree well with those obtained when the image potential is calculated for an actual extended Si waveguide, thus supporting our adoption of the PEC approximation.

% --- Figure 2 --------------------------------------------

% ---------------------------------------------------------
\subsection{Trajectory-integrated excitation probability}
\label{sec_uniform_field}

We are interested in the total excitation probability $P(\omega)$ obtained by integrating the loss probability per unit of path length $\Gamma(z,\omega)$ along the electron trajectory, under the assumption that a parallel e-beam configuration yields an accurate description at each value of the electron--waveguide separation $z$ for glancing incidence. More precisely, $P(\omega)=2\int_0^\infty dy\,d\Gamma[z(y),\omega]/dy$, where we have substituted $b$ by the varying separation $z(y)$ and used the fact that the trajectory is symmetric with respect to the turning point $(y=0,z=b)$. A more practical expression is then obtained by changing the integration variable from $y$ to $z$, so we obtain
\begin{align} \label{eels_oblique} %--
P(\omega)=\sqrt{2\me v^2\gamma}\int_b^\infty dz\;\frac{\Gamma(z,\omega)}{\sqrt{U(b)-U(z)}},
\end{align}
where $U(z)$ is given by Eq.~(\ref{gofz}) and the lower integration limit is defined by the turning point $z=b$.

% ---------------------------------------------------------
\subsection{Uniform external DC field}
\label{sec_uniform_field}

We now consider the system schematically shown in Fig.~\ref{Fig2}(a), consisting of an infinitely extended planar Si waveguide of thickness $h$ and an e-beam moving towards the waveguide with in-plane velocity $\vb\parallel\yy$, incident with a grazing angle $\theta\ll1$. A planar electrode is placed at a distance $d$ above the waveguide, held at a potential difference $V_0$, such that a uniform DC repulsive field $E^{\rm DC}=V_0/d$ is maintained, corresponding to an electron potential energy
\begin{align}
\label{eq_potential_DC}
U^{\rm DC}(z)=-eE^{\rm DC}z.
\end{align}
Incidentally, calculating the image and external forces from the potentials in Eqs.~(\ref{eq_potential_image_approx}) and (\ref{eq_potential_DC}), and plugging the result in Eq.~(\ref{bthreshold}), we obtain the turning-point threshold $b_{\rm th}=e/\sqrt{4\gamma E^{\rm DC}}$. In general, the electron trajectory is parabolic and solely determined by the field $E^{\rm DC}$ and the turning-point distance $b$. The latter is determined by the initial electron--waveguide distance $z_0$ and angle of incidence $\theta$ at the point of electron entrance in the capacitor region [leftmost part of Fig.~\ref{Fig2}(a)]. For simplicity, we take $d$ and $z_0$ to be larger than the penetration depth of the waveguide modes into the vacuum, such that they are not influenced by the top gate, and the electron accumulates a complete interaction with the mode profiles.

In Fig.~\ref{Fig2}(b), we show the spectral dependence of $P(\omega)$ as calculated from Eq.~(\ref{eels_oblique}) for different values of the turning point $b$, considering parabolic trajectories determined by a DC field $E^{\rm DC}=10^8\,$V/m and plugging the EELS probability given by Eq.~(\ref{eels_a_plane}) for the planar waveguide. This figure reveals the excitation of multiple modes with amplitudes that increase as $b$ decreases, including undesired high-energy interband transitions, which are only observed at low $b$. To quantify the photon yield per electron, we define the frequency-integrated EELS probability
\begin{align}
\label{eq_Gamma_f}
N(\omega_{\rm c}) = \int_0^{\omega_{\rm c}} d\omega \, P(\omega),
\end{align}
which depends on the frequency cutoff $\omega_{\rm c}$. Figure~\ref{Fig2}(c) shows that $N(\omega_{\rm c})$ displays an initial increase over the waveguided mode region, followed by a saturation for large $b$ (i.e., when interband transitions are negligible) and a more sustained growth for small $b$. A clearer quantification of the weight carried by waveguided modes in the total loss probability is provided by the normalized quantity $N(\omega_{\rm c})/N(\infty)$, plotted in Fig.~\ref{Fig2}(d) and showing that for the largest $b$'s under consideration most of the losses originate in the excitation of waveguided modes, thus anticipating the formation of a coherent state made of entangled photon--electron components. As we discuss below, the post-interaction state of the waveguide modes is given by a Poissonian distribution of Fock states, while $P(\omega)$ and $N(\omega)$ must be understood as average values per electron.

% --- Figure 3 --------------------------------------------
 \begin{figure*}[htbp]
\centering
\includegraphics[width=0.8\linewidth]{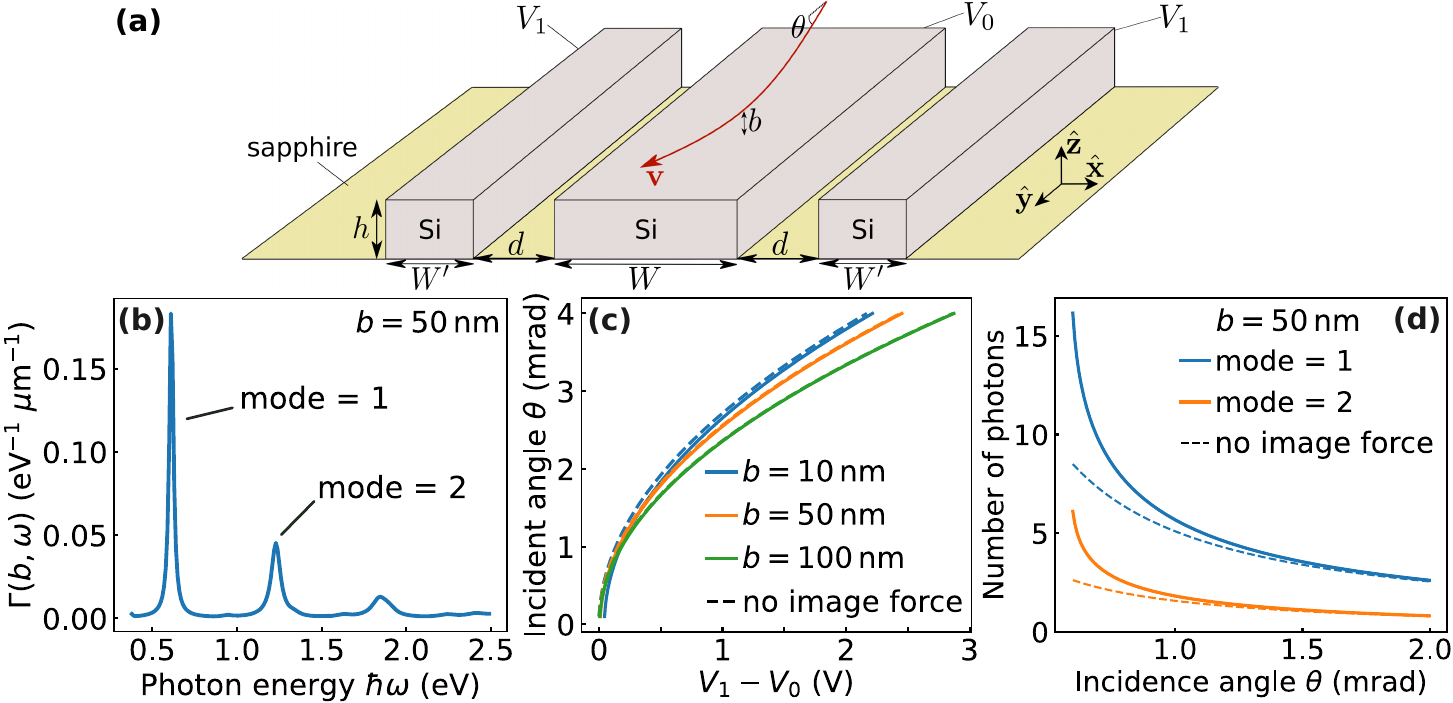}
\caption{\textbf{Electron coupling to waveguide modes in a biased waveguide geometry}. 
\textbf{(a)} We consider a conductive Si central waveguide (voltage $V_0$) flanked by two Si electrodes (voltage $V_1$) supported on a sapphire substrate. The scheme shows the geometrical parameters of the system ($W=500$~nm, $W'=250$~nm, $d=100$~nm, and $h=300$~nm in our calculations). We consider 100~keV electrons moving in the $y-z$ plane of specular symmetry with velocity $v$ parallel to the waveguide. The minimum electron--waveguide distance $b$ is determined by the initial incidence angle $\theta\ll1$.
\textbf{(b)} EELS probability per unit of path length for a fixed electron--waveguide distance of 50~nm.
\textbf{(c)} Relation between incidence angle $\theta$ and bias voltage $V_1-V_0$ for several fixed values of the minimum distance $b$ [Eq.~(\ref{eq_bmin})] with (solid curves) and without (dashed curves) including image-force corrections in the electron trajectory.
\textbf{(d)} Number of photons excited per electron as a function of incidence angle for modes 1 and 2 in (b) with (solid curves) and without (dashed curves) image-force corrections for $b=50$~nm.
The sapphire permittivity is set to 9 for DC potential calculations and 3.13 for EELS in the optical range.
}
\label{Fig3}
\end{figure*}

% ---------------------------------------------------------
\subsection{Spatially varying external DC field}
\label{sec_potential3nanowires}

In a more practical configuration, we consider a 1D waveguide, such that excited modes can be delivered to a desired position in an optical system. In addition, we introduce an in-plane gating configuration as depicted in Fig.~\ref{Fig3}(a). To analyze this system, we use the same procedure as in Sec.~\ref {sec_uniform_field}, so that the trajectory-integrated loss and photon-emission probabilities are given by Eqs.~(\ref{eels_oblique}) and (\ref{eq_Gamma_f}), with the potential energy $U(z)$ still written as in Eq.~(\ref{gofz}). For simplicity, we adopt the PEC approximation to calculate the image potential [Eq.~(\ref{eq_potential_image_approx})], which is reasonable because it mainly acts at short electron--waveguide distances, smaller than the width of the waveguide, which can therefore be assimilated to an infinite plane. However, we need to consider the actual waveguide geometry to calculate the external DC field and the loss probability.

We compute the EELS probability $\Gamma(b,\omega)$ per unit of path length for the 1D waveguide using the boundary-element method\cite{paper040} (BEM) as a function of electron--surface distance $b$ and loss frequency $\omega$. In particular, we use an implementation of BEM adapted for systems with translational symmetry along one spatial direction.\cite{paper040} We consider the waveguide to be supported on a sapphire planar surface, for which the optical permittivity is approximated as a constant value of 3.13 in the optical energy range of the excited modes under consideration. For Si, we use the same permittivity as in Sec.~\ref{sec_parallel_trajectory}.\cite{AS1983} We ignore the effect of the side gates on the EELS probability, as their distance to the waveguide is larger than the penetration of the mode fields in vacuum (see below). More details on the calculation are offered in Appendix~\ref{section_bem2D}, where we also fit the $z$ dependence of the resulting probability to an exponentially decaying function. Figure~\ref{Fig3}(b) shows the EELS probability per unit of path length for a 100~keV electron passing at a distance $b=50$~nm from a sapphire-supported Si waveguide of width $W = 500$~nm and height $h=300$~nm. The spectrum allows us to identify two main modes (labeled 1 and 2) at energies $\hbar\omega_1 = 0.61$~eV and $\hbar\omega_2 = 1.23$~eV. In Fig.~\ref{Fig3}, we concentrate on the excitation of these two modes, whose associated numbers of excited photons are obtained by integrating over their respective spectral peaks.

To obtain the external DC potential $U^{\rm DC}(x,z)$ produced in the configuration of Fig.~\ref{Fig3}(a), we treat the waveguide and the flanking electrodes as PECs and parametrize their surfaces through boundary elements to self-consistently calculate the resulting surface charge and 3D potential distributions (see details in Appendix~\ref{gating1D}). The sapphire substrate is described by a static permittivity of 9. Notice that the potential also depends on the in-plane transverse coordinate $x$, although for concreteness we consider the electron to be evolving in the central $x=0$ plane. Incidentally, compared to a vertical gating configuration (e.g., with the waveguide placed above a planar bottom gate), the lateral gating configuration appears to be more amenable to conventional nanofabrication methods such as e-beam lithography and focused-ion-beam milling, and it produces more repulsive DC fields for comparable applied voltages (see Appendix~\ref{gating1Dbis}).

Far from the waveguides ($z\to\infty$), the potential energy vanishes [$U(\infty)=0$] and the electron moves with a glancing angle $\theta$ relative to the waveguide (i.e., $dz/dy=\tan\theta\ll1$). Then, Eq.~(\ref{eq_dy_electron_aux}) leads to
\begin{align}
\label{eq_bmin}
\tan^2\theta=\frac{2U(b)}{\me v^2\gamma}.
\end{align}
Note that the right-hand side of this expression depends on the applied voltage difference $V_1-V_0$ between the side gates and the waveguide. Figure~\ref{Fig3}(c) shows the relation between the incidence angle $\theta$ and the bias voltage for several fixed values of $b$ and 100~keV electrons. The solid curves are calculated with the inclusion of image-force corrections, while dashed curves disregard it. The two sets of curves nearly overlap for large $b$, indicating that the image attraction plays only a small role in defining the minimum electron--waveguide distance.

To obtain the total number of photons created, we plug the DC potential energy and EELS probability into Eq.~(\ref{eels_oblique}) and integrate the result over the spectral width of each of the two lowest-order modes under consideration. Figure~\ref{Fig3}(d) shows the total number of photons excited per electron as a function of the incidence angle $\theta$ for the first and second modes (blue and orange solid curves, respectively), for $b=50$~nm. The first mode reaches a sizeable yield of $\sim15$ photons per electron, while the second mode reaches $\sim5$ photons per electron (both for $\theta \sim 0.6$~mrad, corresponding to $V_1-V_0\sim45$~mV). For comparison, the dashed curves show the photon yield obtained when the image force is neglected. The image force enhances the electron--waveguide interaction for each fixed value of $\theta$ because the potential energy becomes smoother and the electron ends up spending more time near the waveguide (see Appendix~\ref{gating1D}), resulting in a higher number of photons for both modes. The photon yield decreases with increasing incidence angle because a higher voltage is then required to maintain the same turning point, which repels the electron faster from the waveguide and reduces the interaction time. Reducing $b$ would systematically increase the photon yield, but the condition $b>b_{\rm th}$ must be satisfied to avoid electron--waveguide collisions and prevent incoherent excitation of interband transitions in Si.
 
% =========================================================
\section{Conclusions}

We have shown that free electrons can couple efficiently to guided optical modes of dielectric waveguides when their trajectories are engineered to remain close to the photonic structure over an extended interaction length. Starting from the reference case of an electron moving parallel to a planar silicon waveguide, we found that the EELS spectrum directly reflects the guided-mode dispersion and that the excitation probability increases rapidly as the electron–waveguide separation is reduced. This regime allows selective coupling to different TE and TM branches and provides a direct route to waveguided photon generation by focused e-beams.

We then demonstrated that electrostatically deflected trajectories can strongly enhance the interaction. In particular, grazing trajectories with a well-defined turning point near the waveguide combine a small electron–mode separation with a long dwell time in the near field, leading to integrated excitation probabilities of order unity and, for realistic parameters, to multiphoton emission into guided modes. The coupling strength and spectral distribution can be tuned through the incidence angle, electron energy, and applied electrostatic bias.

A central practical issue is the image interaction between the electron and the dielectric structure. This self-interaction can pull the electron into the surface and therefore sets a threshold closest-approach distance below which stable propagation is not possible without external compensation. We have included this effect in the trajectory analysis and shown that appropriately designed electrostatic fields can overcome it, enabling controlled grazing trajectories compatible with strong waveguide coupling.

An important clarification is that the present analysis evaluates photon generation within a classical EELS framework. The calculated excitation probabilities should therefore be interpreted as mean photon numbers for the corresponding waveguide modes, rather than as deterministic photon-generation events. In a quantum description, each mode is populated by a distribution of Fock states, with Poissonian statistics when the electron acts as a classical source. The electron energy-loss spectrum then records the number of emitted quanta: an electron that has lost an energy equal to the sum of photon energies in the occupied modes is correlated with the corresponding multimode photonic Fock state \cite{paper455}. Consequently, the outgoing electron and the waveguided field form an entangled state in the photon-number basis, with correlations between the discrete electron energy losses and the occupation numbers of each guided mode. In practice, the measured signal corresponds to an incoherent mixture of such electron--photon-number correlated states. 

We remark that the analyzed gated waveguide geometries can generate the required electrostatic landscapes in an integrated platform. These structures provide a practical mechanism for steering the electron trajectory while maintaining optical confinement in the waveguide. The results indicate that electrically controlled electron paths can be used to enhance and tune free-electron coupling to guided photonic modes, opening a route toward deterministic or near-deterministic generation of waveguided photons and multiphoton Fock states in electron-driven integrated nanophotonic devices.

\acknowledgments
This work has been supported in part by the European Research Council (101141220-QUEFES), the Spanish MICIU (PID2024-157421NB-I00 and Severo Ochoa CEX2024-001490-S), and the CERCA Program.

\section*{DATA AVAILABILITY}

The data supporting the findings of this study are available within the article.

\appendix

% --- Figure 4 --------------------------------------------
\begin{figure}[htbp]
\centering
\includegraphics[width=1.0\linewidth]{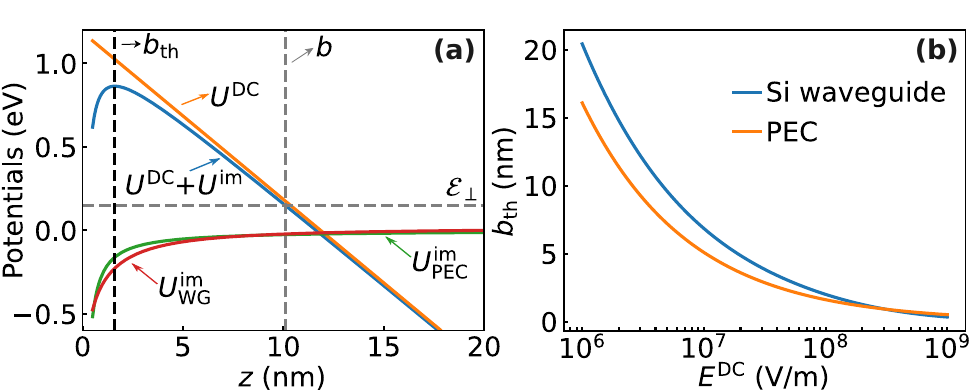}
\caption{\textbf{Effect of the image potential model on the threshold electron--waveguide distance}. \textbf{(a)}~Out-of-plane electron potential energy $U(z)=U^{\rm DC}(z)+U^{\rm im}(z)$ and its DC and image components for $E^{\rm DC}=10^8\,$V/m. We indicate the turning-point distance $b\approx10~$nm for a specific normal electron energy $E_\perp$ (horizontal dashed line), as determined by the intersection with the potential. For $b$ below the threshold value $b_{\rm th}$, the trajectory impacts the waveguide surface at $z=0$. We compare the image potential for a PEC and a Si waveguide of 200~nm thickness.
\textbf{(b)}~Dependence of the threshold distance $b_{\rm th}$ on the external DC field $E^{\rm DC}$, calculated when the image potential is either taken for a PEC surface [Eq.~(\ref{eq_potential_image_approx})] or calculated from Eq.~(\ref{Fimgeneral}) for a Si waveguide of 200~nm thickness. The electron velocity is set to $v=0.7\,c$ in both panels.}
\label{Fig4}
\end{figure}

% =========================================================
\section{Image force for a moving electron}
\label{image_force}

We adopt the expression derived in Ref.~\citenum{paper455} for the image force experienced by an electron moving at a distance $z$ parallel to a planar surface characterized by Fresnel's reflection coefficients $\rp(\kpar,\omega)$ and $\rs(\kpar,\omega)$,
\begin{align} \label{Fimgeneral} %--
F^{\rm im}_z(z) &= - \frac{2e^2}{\pi v^2\gamma} \int_0^\infty\omega d\omega\int_0^\infty d k_x \; \e^{-2\kappa z} \\ 
&\times\Big[(1-k^2/\kpar^2)\,\Ree\{\rp\}+(k_x^2/\kpar^2)(v/c)^2\,\Ree\{\rs\}\Big], \nonumber
\end{align}
where $\kpar=\sqrt{(\omega/v)^2+k_x^2}$ and $\kappa=\sqrt{(\omega/v\gamma)^2+k_x^2}$. This expression applies to any planar homogeneous surface. For a PEC, we have $\rp\to 1$ and $\rs\to-1$, and the integrals in Eq.~(\ref{Fimgeneral}) can be performed analytically, leading to $F^{\rm im}_z(z)=-e^2/4\gamma z^2$, so the image potential reduces to Eq.~(\ref{eq_potential_image_approx}). This result is consistent with a previous quantum-fluctuation analysis.\cite{F93}

In this work, we assimilate the waveguide to an extended planar surface and model Si as a PEC. To support this approximation, we compare the so-calculated image potential energy to the result obtained from Eq.~(\ref{Fimgeneral}) for a Si waveguide of 200~nm thickness. The latter enters Eq.~(\ref{Fimgeneral}) through the Fresnel reflection coefficients, which are in turn obtained as described in Sec.~\ref{sec_parallel_trajectory}. Both models agree, except for small electron--waveguide separations [Fig.~\ref{Fig4}(a)], which should be avoided in practice to avert the excitation of interband transitions in Si. We also compare the threshold for the electron--surface distance calculated either in the PEC-model [Eq.~(\ref{bthreshold})] or in the actual Si waveguide, showing excellent mutual agreement and further justifying our use of the PEC image potential [Fig.~\ref{Fig4}(b)].

% --- Figure 5 ---------------------------------------------
\begin{figure}[htbp]
\centering
\includegraphics[width=1.0\linewidth]{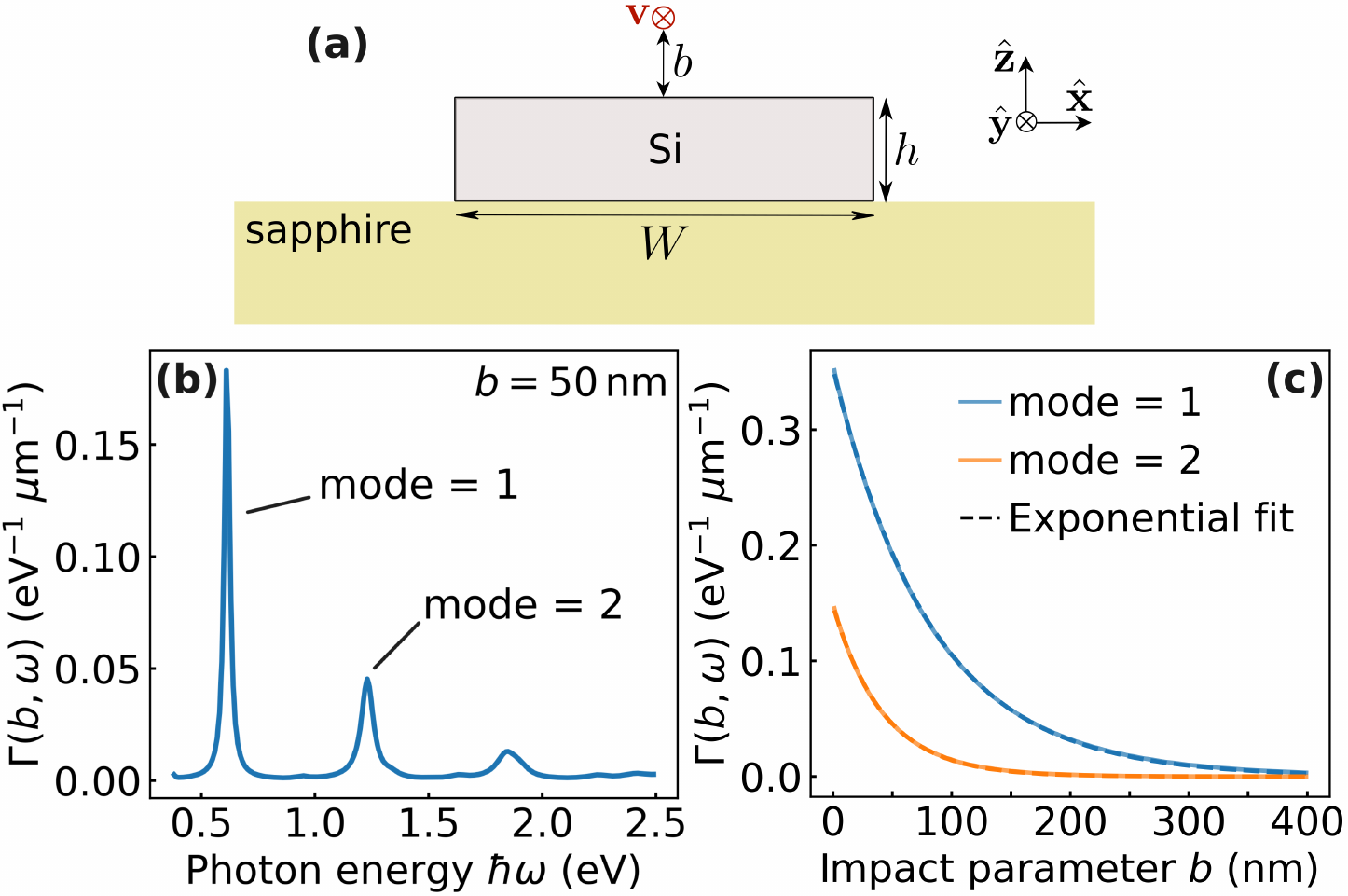}
\caption{\textbf{Calculation of the EELS probability and its exponential decay away from a 1D waveguide}.
\textbf{(a)} Transverse cross-section of the geometry under consideration, including an electron (red dot) moving with velocity $v$ at a distance $b$ from the sapphire-supported Si waveguide. 
\textbf{(b)} EELS probability per unit of path length as obtained from BEM for $b = 50$~nm, $W=500$~nm, $h=300$~nm, and 100~keV electrons.
\textbf{(c)} Dependence of the peak probability for modes 1 and 2 on $b$ (solid curves) and exponential fits (dashed curves).}
\label{Fig5}
\end{figure}

% =========================================================
\section{Calculation of the EELS probability in the 1D waveguide}
\label{section_bem2D}

We calculate the EELS probability per unit of path length $\Gamma(b,\omega)$ using a BEM implementation for structures with translational invariance along one direction.\cite{paper040} The simulated geometry is depicted in Fig.~\ref{Fig5}(a), where an electron (red dot) moves with velocity $v$ at a distance $b$ above and parallel to a Si waveguide (width $W$, height $h$). The latter is supported on a sapphire substrate (optical permittivity of 3.13). Figure~\ref{Fig5}(b) shows an example of a calculated EELS spectrum. As shown in Fig.~\ref{Fig5}(c), we find that the $b$ dependence of the EELS probability is well approximated by an exponential decay as $\Gamma(b,\omega)\propto\ee^{-2\kappa b}$, with the $\kappa$ parameter maintained as a constant within the spectral region associated with each waveguided mode.

% --- Figure 6 --------------------------------------------
\begin{figure}[htbp]
\centering
\includegraphics[width=\linewidth]{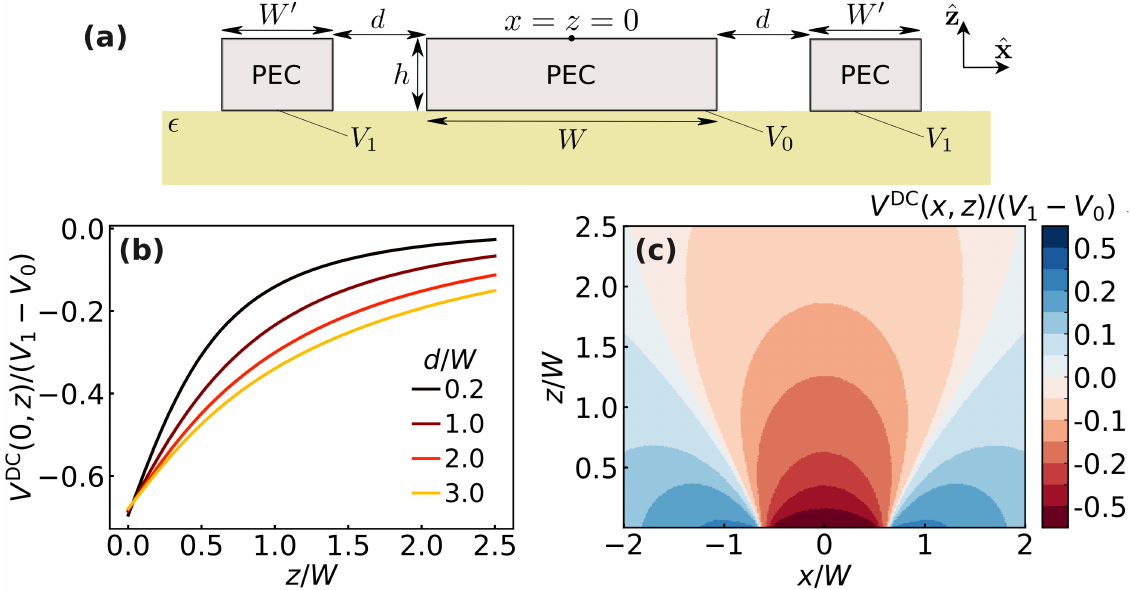}
\caption{\textbf{DC potential in a laterally gated waveguide}.
\textbf{(a)}~Scheme of the structure under consideration, corresponding to Fig.~\ref{Fig3}(a) with Si described as a PEC and the dielectric substrate of DC permittivity $\epsilon$.
\textbf{(b)}~Normalized DC potential $V^{\rm DC}(x=0,z)/(V_1-V_0)$ as a function of $z/W$ for different values of the $d/W$ ratio (solid curves) compared to linear fits at low $z$ (dashed lines) for $W'=h=0.5\,W$.
\textbf{(c)}~Full $(x,z)$ dependence of the normalized DC potential for $W'=h=0.5\,W$ and $d=0.2\,W$.}
\label{Fig6}
\end{figure}

% =========================================================
\section{Calculation of the DC potential in a laterally gated 1D waveguide}
\label{gating1D}

Here, we compute the DC potential distribution $V^{\rm DC}(x,z)$ from which we obtain the electron potential energy $U^{\rm DC}=-eV^{\rm DC}$ used in Sec.~\ref{sec_potential3nanowires}, which is generated by three linear PEC wires (the waveguide at voltage $V_0$ and two side gates at $V_1$) placed on the surface ($z=0$ plane) of a dielectric of permittivity $\epsilon$, as shown in Fig.~\ref{Fig6}(a). All wires have the same height $h$, the width of the central one is $W$, and the side gates have widths $W'$. The lateral spacing between neighboring wires is $d$. Charges are induced on the surface of the PEC wires, such that we can express the potential distribution in the vacuum region above the dielectric as
\begin{align} \label{eq_potential_aux_3n} %--
V^{\rm DC}(x,z)&=\int_C ds\int dy \;\sigma(s) \\
\times\bigg[&\dfrac{1}{\sqrt{|(x-x_s,z-z_s)|^2+y^2}} \nonumber\\
+&\dfrac{g_\epsilon}{\sqrt{|(x-x_s,z+z_s)|^2+y^2}} \bigg]. \nonumber
\end{align}
Here, we integrate the surface charge density $\sigma(s)$ over the transverse contour $C$ of the metallic wires, parametrized through the distance $s$. The first fraction in Eq.~(\ref{eq_potential_aux_3n}) represents the direct Coulomb potential, while the second term accounts for the image charge proportional to $g_\epsilon=(1-\epsilon)/(1+\epsilon)$ in the dielectric medium. We also introduce an integral over the direction of translational symmetry $y$. The latter can be performed analytically, while the $s$ integral can be reduced to the right-hand side of the structure ($x_s>0$) by leveraging its mirror symmetry with respect to the $x=0$ plane. Combining these elements, Eq.~(\ref{eq_potential_aux_3n}) leads to
\begin{align} \label{eq_final_potential_3nanowires} %--
V^{\rm DC}(x,z)&=\int_{C,x_s>0} ds \;M(x,z,s) \;\sigma(s)
\end{align}
with
\begin{align} \label{Mxz} %--
M(x,z,&s) \\
=2\Big\{&\log\big[|(x-x_s,z-z_s)||(x+x_s,z-z_s)|\big] \nonumber\\
+g_\epsilon&\log\big[|(x-x_s,z+z_s)||(x+x_s,z+z_s)|\big]\Big\}, \nonumber
\end{align}
where the integral is restricted to the $x_s>0$ region of the PEC wire contours. We then approximate the structure by selecting a finite set equally spaced contour points $s_j$, such that Eq.~(\ref{eq_final_potential_3nanowires}) can be recast into
\begin{align} \label{VMsigma} %--
V^{\rm DC}(x,z)=\Delta_s\sum_j M(x,z,s_j)\,\sigma_j,
\end{align}
where $\Delta_s$ is the spacing between consecutive points. The self-consistent charges $\sigma_j\equiv\sigma(s_j)$ are then determined by imposing the values $V^{\rm DC}_j=V_0$ and $V^{\rm DC}_j=V_1$ for the DC potential at the contour points $s_j$ on the central and side gates, respectively, so we obtain the linear set of equations $V^{DC}_j=\Delta_s\sum_{j'} M(x_j,z_j,s_{j'})\,\sigma_{j'}$. Once the charges $\sigma_j$ are determined, we use Eq.~(\ref{VMsigma}) to compute the potential at any point $(x,z)$ in the vacuum region outside the wires and the dielectric.

In Fig.~\ref{Fig6}(b), we show the potential distribution normalized to the bias voltage $V_1-V_0$ for different values of the $d/W$ ratio, taking $\epsilon=9$, as corresponding to the DC permittivity of sapphire. The potential exhibits a nearly linear decay close to the wire surface, followed by a smoother profile at larger distances. For the dimensions under consideration in Fig.~\ref{Fig3}, the extension of the evanescent mode field outside the waveguide lies within the region of linear potential decay. Interestingly, the slope (i.e., $E^{\rm DC}$) in the linear region is larger for smaller separations $d$, therefore leading to stronger repulsive fields for a given value of the potential bias. The full $(x,z)$ dependence of the voltage is shown in Fig.~\ref{Fig6}(c) for the smaller value of the ratio $d/W=0.2$ under consideration, showing a relatively extended region above the central waveguide in which the potential is repulsive for the electron.

% =========================================================
\section{Calculation of the DC potential in a vertically gated 1D waveguide}
\label{gating1Dbis}

\begin{figure}[htbp]
\centering
\includegraphics[width=\linewidth]{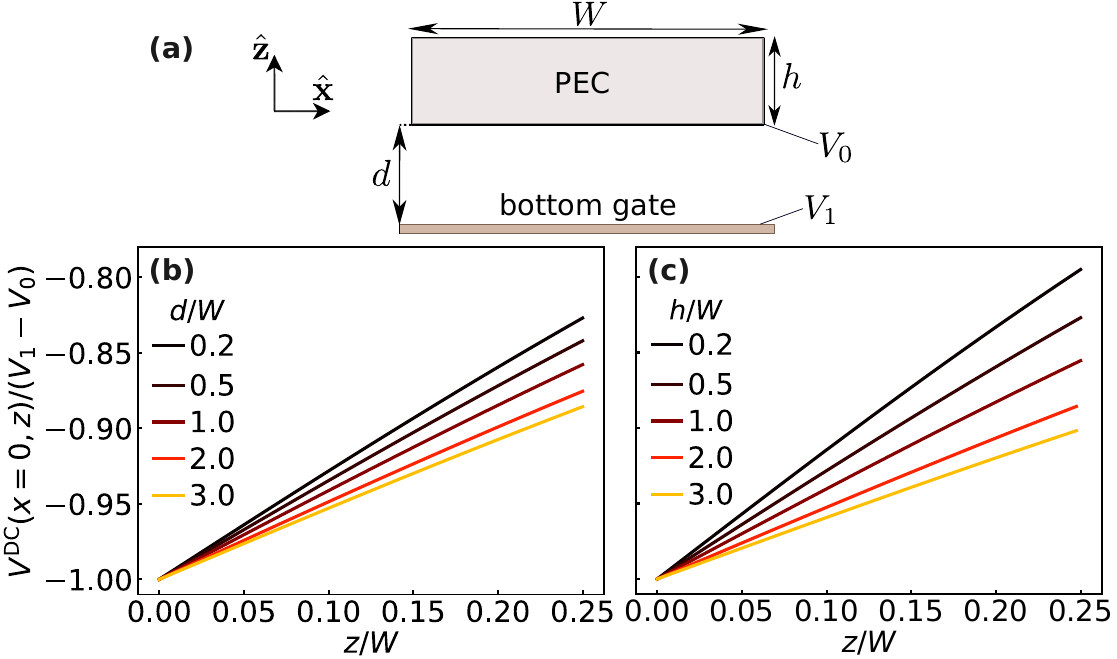}
\caption{\textbf{DC potential in a vertically gated waveguide}.
\textbf{(a)}~Scheme of the structure under consideration, consisting of a self-standing PEC wire (width $W$, thickness $h$, voltage $V_0$) separated by a distance $d$ from a bottom PEC gate (voltage $V_1$). 
\textbf{(b)}~Normalized DC potential $V^{\rm DC}(x=0,z)/(V_1-V_0)$ as a function of $z/W$ for different values of the $d/W$ ratio and fixed $h/W=0.5$.
\textbf{(c)}~Same as (b) for different values of $h/W$ and fixed $d/W=0.2$.}
\label{Fig7}
\end{figure}

As an alternative configuration to lateral gating, we consider the structure depicted in Fig.~\ref{Fig7}(a), where the waveguide stands at a distance $d$ from a planar bottom gate. Both the waveguide and the bottom gate are treated as PECs. We calculate the potential distribution in this geometry using the same methods as in Appendix~\ref{gating1D}, but with $g_\epsilon=-1$ to account for the image of the waveguide created by the bottom gate. We also introduce a voltage bias $V_1-V_0$ between the bottom gate and the waveguide. Unfortunately, as shown in Fig.~\ref{Fig7}(b,c), the slope of the potential is much weaker in this configuration than in the lateral gating structure, and therefore, we only present results for photon generation for the latter.

% --- Figure 7 --------------------------------------------

%\bibliography{../../../bibtex/refsL.bib}


\end{document}